\documentclass[12pt,preprint]{aastex}

\slugcomment{}

\shorttitle{Mauna Kea Sky Transparency}
\shortauthors{Steinbring, Cuillandre \& Magnier}

\input epsf
\def\plotone#1{\centering \leavevmode
\epsfxsize=1.0\columnwidth \epsfbox{#1}}

\def\plotonesmall#1{\centering \leavevmode
\epsfxsize=0.75\columnwidth \epsfbox{#1}}

\begin{document}

\title{Mauna Kea Sky Transparency from CFHT SkyProbe Data}

\author{Eric Steinbring\altaffilmark{1}, Jean-Charles Cuillandre\altaffilmark{2} and Eugene Magnier\altaffilmark{3}}

\altaffiltext{1}{Herzberg Institute of Astrophysics, National Research
Council Canada, Victoria, BC V9E 2E7, Canada}
\altaffiltext{2}{Canada-France-Hawaii Telescope, Kamuela, HI 96743}
\altaffiltext{3}{University of Hawaii, Institute for Astronomy, Honolulu, HI 96822}
 
\begin{abstract}
Nighttime sky transparency statistics on Mauna Kea are reported based on data from
 the Canada-France-Hawaii Telescope SkyProbe monitor. We focus on the period
beginning with the start of MegaCam wide-field optical imager operations in 2003, and
continuing for almost three years. Skies were clear enough to observe 
on 76\% of those nights; attenuations were less than 0.2 magnitudes up to 60\% of the time. An empirical model of cloud attenuation and duration is presented allowing us to further characterize the photometric conditions. This is a good fit to the SkyProbe data, and indicates that Mauna Kea skies are truly photometric (without cloud) an average of 56\% of the time, with moderate seasonal variation. Continuous monitoring of transparency during the night is necessary to overcome fluctuations in attenuation due to thin cloud.
\end{abstract}

\section{Introduction}\label{introduction}

Sky transparency is a fundamental parameter governing the operation
of any optical observatory.  Best telescope performance comes
when skies are photometric - cloud-free - and in general, efficiency at a given
site depends directly on the fraction of time that skies are clear enough to observe. As more telescopes employ adaptive optics (AO) facilities
employing lasers, even thin cloud can have a significant impact 
on telescope operations. Attenuation of the beam from cirrus will not only reduce laser return power and
affect AO wavefront-sensor performance, but
reflected Rayleigh-backscatter light may be of concern to other nearby optical observatories.

Several sources of sky clarity measurements are available for Mauna Kea. The longest historical
records date back to the 1970s, consisting of observers' nightly notes on sky conditions 
in the logbooks of the University of Hawaii 88-inch, United Kingdom Infrared Telescope (UKIRT), and 
the Canada-France-Hawaii Telescope (CFHT). Quantitative measurements have been made in 
``campaign mode'' \citep{McCord1979, Krisciunas1987}. But only in the last few years has
there existed dedicated instruments for measuring sky clarity.
Two of these are the Continuous Camera (CONCAM) operated by the {\it Night Sky Live} network since 2005 \citep{Shamir2005} and
the Thirty Meter Telescope (TMT) All Sky Camera (ASCA) installed in 2006.
Another is the CFHT SkyProbe. The CONCAM and TMT ASCA are all-sky imagers, and as such are best used 
to provide a panoramic view of cloud cover above the summit.
SkyProbe is specifically designed to
measure sky transparency as an input to telescope data reduction \citep{Cuillandre2002}.

SkyProbe is fixed to the side of the 
CFHT. While the dome is open, once every minute it images bright stars 
in a 35 square-degree field overlapping the telescope field of view, performs automated photometry, and returns a measurement
 of attenuation corrected for zenith angle.  
It derives an absolute zeropoint based on the Hipparcos' Tycho catalog, which provides a solid 
photometric reference; uncertainty is on the order of 0.01 mag. An excellent feature of this design is that this measurement is in the direction
of the telescope, and as such can be directly compared to a concurrent zeropoint measurement
of another instrument, for example the MegaCam imager.
The consistency of the SkyProbe data allows the ensemble of measurements taken during a night, or over more extended time periods, to be used to estimate general properties of the atmosphere. Nearby weather measurements from the CFHT meteorological tower can be used to check for consistency. 
SkyProbe does not take data at all times though, as it is shut off when the dome is closed.

One utility of SkyProbe, helpful for the observer in pruning data of poor quality, is the detection of clouds. Thick clouds, say 0.5 magnitudes in extinction or more, are easily recognized by sharp fluctuations in attenuation. Less obvious are thin clouds, especially as their attenuations approach that of the atmosphere itself. To elucidate this, it is useful to separate attenuation into air and cloud components as
$$A=A_{\rm air}+A_{\rm cloud}=A_{\rm air, 1}Z+A_{\rm cloud,1}T, \eqno(1)$$
where cloud thickness $T$ is analogous to airmass $Z$. Like $Z$, it parameterizes the bulk quantity of absorber without making any assumptions about its distribution along the beam. Constants $A_{\rm air,1}$ and $A_{\rm cloud,1}$ indicate attenuations for $Z=1$ and $T=1$. Separated in this way, it can be seen that if conditions are usually photometric, 
$$A_{\rm air,1}={\hat A}, \eqno(2)$$
where ${\hat A}$ indicates the mode of $A$, and 
$$A_{\rm cloud,1}={\bar A}-{\hat A}, \eqno(3)$$
where ${\bar A}$ is the mean. So clouds are detectable if their attenuations are greater than ${\hat A}$ or when $T$ varies on timescales shorter than $Z$ - or both.

Under photometric skies, astronomers often fix their photometry to a zeropoint assuming an airmass $Z=1$ at zenith for the observatory, although that is not strictly true. The density of air fluctuates with barometric pressure and temperature, and the concentrations of water vapour and other atmospheric constituents such as aerosols \citep[see, e.g.][and references therein]{Chambers2005}. To first order, however, airmass will increase linearly with barometric pressure $p$ and relative humdity $\rho$, 
$$Z=p \rho/({\bar p}{\bar \rho}). \eqno(4)$$
A complication is that $\rho$ measured on the ground does not necessarily correlate with the bulk of the atmosphere. One extreme case is fog. But barometric pressure is generally stable during a night. 
So a plausible empirical model - requiring input only from SkyProbe - is that ${\hat A}$, calculated nightly, tracks the observed atmospheric attenuation.
And an even simpler model avoids the details of fluctuations in airmass entirely. This assumes that highs in barometric pressure (or any other atmospheric parameter) will be balanced by corresponding lows, giving a Gaussian distribution with a long term average of
$$\bar{Z}={1\over{\sqrt{2\pi}\sigma}}\int_0^\infty{\exp{{\Big(}-{\textstyle (Z-1)^2\over{2\sigma^2}}{\Big)}}}\,{\rm d}Z=1, \eqno(5)$$
where $\sigma$ is the standard deviation in nightly $\hat A$, expressed as a percent.

When clouds are present, they may go unnoticed if they are thin, persistent and of uniform thickness. One possible means of detecting thin cloud is motivated by the fractal properties of cirrus (and stratus) clouds, in that their density and lifetime are correlated over many orders of magnitude, governed by a tight power law \citep{Ivanova2000, Ivanova2002}. If so, the number of SkyProbe samples of attenuation $A_{\rm cloud}$ - the distribution of attenuations - should follow
$$N(A_{\rm cloud})\propto\exp(-A_{\rm cloud}/\alpha A_{\rm cloud,1}), \eqno(6)$$
where $\alpha$ is a constant and the thickness of clouds is given by $T=A_{\rm cloud}/A_{\rm cloud,1}$. For uniformly timed samples, the number of samples of a given attenuation should then be directly proportional to the duration of a cloud, $\delta$, and hence
$$\delta=\delta_0\exp(-T/\alpha), \eqno(7)$$
where $\delta_0=f_{\rm clear}\,t_{\rm obs}$ is the duration of clear skies during an observation lasting $t_{\rm obs}$, and $f_{\rm clear}$ is the fraction of time that is clear. This implicitly assumes that samples are taken whatever the cloudiness, that is, without bias towards pointing the telescope at clearer parts of the sky, for example. That seems reasonable for conditions so good that clouds are hard to detect. We will simply assume for now that samples are unbiased, and see if that still holds as clouds become thicker.
Because the average thickness of a cloud is defined to be 
$$\bar{T}=\alpha\log(\delta_0/{\bar{\delta}})=1,\eqno(8)$$
and since
$$\bar{\delta}/\delta_0={1\over{T_{\rm max}}}\int_0^{T_{\rm max}}\exp{{\Big(}-{T\over{\alpha}}{\Big)}}\,{\rm d}T\approx{\exp(-1/\alpha)}\eqno(9)$$
for maximal thickness $T_{\rm max}>\alpha$,
the thickness of a cloud would be given by
$$T=1-\alpha\log{(\delta/{\bar \delta})}. \eqno (10)$$
This suggests that an observer, after subtracting attenuation due to air, and knowing only $\alpha$, could at least estimate the possible variation in cloud thickness and attenuation during a night, even if individual clouds themselves are hard to detect.

In Section~\ref{data} we present the SkyProbe data and compare them to contemporaneous MegaCam photometric zeropoint measurements.
Then, in Section~\ref{analysis} we show that our simple models of $Z$ and $T$ are a good match to these 
data, reproducing both the effects of atmospheric attenuation and up to 3 magnitudes of extinction 
by clouds. A summary of the results follows in Section~\ref{summary}.

\section{CFHT Data}\label{data}

SkyProbe $V$-band data were obtained from the period 1 January 2003 through 31 October 2005. Each record runs from 6 PM to 6 AM Hawaii Standard time, with very few nights lost to the instrument being down. The first and last hour of data were deleted to avoid high sky brightness, which can lead to incorrect attenuation measurements due to slight nonlinearity in the SkyProbe detector.
Weather data from the CFHT meteorological tower were obtained for the same period. The fraction of time when either SkyProbe did not take data, or the dome shutter was closed (including due to clouds) was 
31.2\%. It should be noted that the years 2003 through 2005 include some of the worst winter weather on record at CFHT. 
Even so, the mean attenuation from SkyProbe is $0.160\pm0.005$ magnitudes per airmass, which is comparable, and even slightly better than the value of $0.206\pm0.023$ reported by \cite{Krisciunas1987} based on UKIRT data spanning 1980 to 1986.
This result is also in good agreement with observer estimates of the sky-transparency for MegaCam during the period of
overlap with the SkyProbe data. Table~\ref{table_statistics} is a breakdown of the 
average nights per year lost to each instrument due to weather and telescope downtime, along with the remainder with attenuation less than 0.2 magnitudes.

\section{Analysis and Discussion}\label{analysis}

\begin{figure}
\plotone{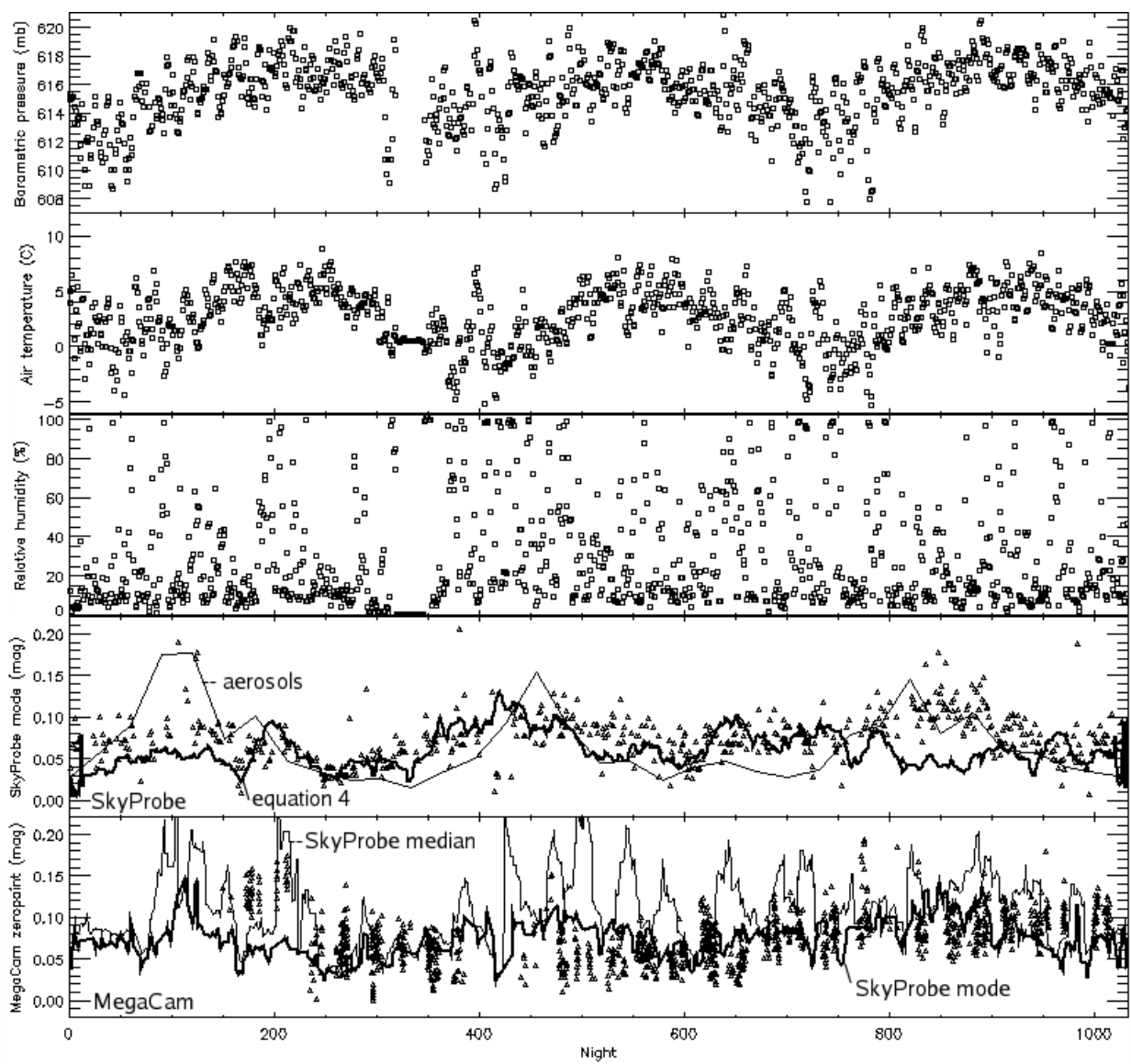}
\caption{Plots of nightly median barometric pressure, temperature, and relative humdity from the CFHT weather tower. Below are nightly modal $V$-band attenuations from SkyProbe and $g$-band zeropoints from MegaCam for the same nights. Atmospheric attenuation is overplotted for SkyProbe assuming equation 4 (thick curve), along with normalized monthly absorption due to aerosols (thin curve) estimated from nephelometer readings at Mauna Loa Observatory. Overplotted for MegaCam are nightly SkyProbe modes (thick curve) and medians (thin curve).}
\label{figure_data}
\end{figure}

Nightly modal SkyProbe attenuations and MegaCam zeropoints are plotted in Figure~\ref{figure_data}, along with
air pressure, temperature, and relative humidity for the CFHT weather tower for the same period. 
Note the strong seasonal variation in air pressure and temperature, with winters clearly evident.
There is no obvious seasonal dependence for relative humidity, apart from increased occurrences of saturation in winter. Modal attenuations from SkyProbe vary slowly from night to night, for the most part, with smooth oscillations over timescales of a month or more. The value of $A_{\rm air, 1}$ inferred is $0.065\pm0.027$, taken to be the global mode of attenuations and the standard deviation of nightly modes over the same period. 

For comparison, equation 4 is overplotted on the SkyProbe data as a thick curve, boxcar smoothed to better show trends over a month. This agrees reasonably well with SkyProbe apart from some deviations, notably an underestimate for about three months in summer 2005 (beginning with night 800). However, there are several sources of atmospheric attenuation unaccounted for in equation 4.  For example, Rayleigh scattering and absorption by ozone have a minimum of about 0.03 magnitudes at 0.55 $\mu$m for the altitude and latitude of Mauna Kea, according to the standard calculation of \cite{Hayes1975}.
Another significant source of atmospheric attenuation is aerosols. Nephelometer measurements of aerosol absorption at nearby Mauna Loa Observatory were downloaded from the National Oceanic and Atmospheric Administration public archives. Averaged monthly absorptions (in the 0.5 $\mu$m channel) are plotted as a thin curve in Figure~\ref{figure_data} for the same period. These have been normalized to a mean attenuation of 0.065 magnitudes, to help show trends relative to equation 4. Variation in aerosol concentration is seen, and one period of elevated levels does correspond to the ``enhancement'' in SkyProbe attenuations in spring/summer 2005 already mentioned. Another rise during the spring of 2003 (beginning approximately on night 50) might be evident in SkyProbe data; a third possibly in spring 2004 (night 450). 
Seasonal variation in the concentration of aerosols has been seen at other observatories, notably in central Europe
\citep{Reimann1992, Pakstiene2001}. There the variation is thought to be the effect of 
higher summer temperatures raising smoke and soot into the upper atmosphere, combined with greater absolute humidity. It is not clear that the same mechanisms would be at work on Mauna Kea, although the seasonal pattern seems similar. One final ``contaminant'' of SkyProbe atmospheric attenuation estimates is evidently thin cloud itself. A few nights even have modal attenuations of up to 0.45 magnitudes.
Further color information might be helpful in discriminating between thin cloud and particulates, and a blue filter (to match the Hipparchos catalog) has recently been installed on SkyProbe \citep{Cuillandre2008}.

The bottom panel of Figure~\ref{figure_data} shows MegaCam zeropoints (subtracted from the minimum zeropoint to indicate attenuation) over the same period. Overplotted are the nightly mode (for $A<0.2$ mag) and median of SkyProbe, the latter highlighting cloudy periods, smoothed to help show trends over a month. Note that MegaCam zeropoints are already
corrected by the observer for atmospheric attenuation, based partly on comparison with SkyProbe. It is possible that at times thin cloud may make this task more difficult, and the result more noisy, although the ``striping'' of zeropoints is due to a known processing problem unrelated to SkyProbe: residual uncorrected chip-to-chip offsets in MegaCam. The value of ${\hat A}$ inferred from MegaCam is $0.074\pm0.031$, in agreement with SkyProbe ($0.065\pm0.027$).

\subsection{Separating Cloud from Air}

\begin{figure}
\plotone{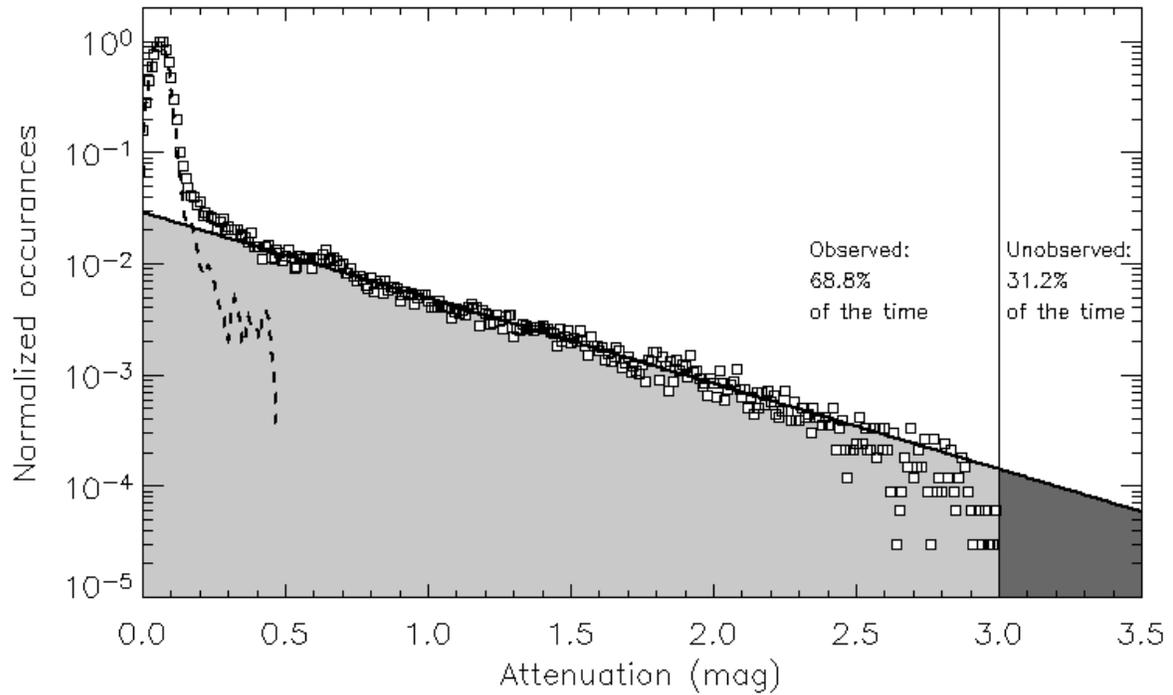}
\caption{A histogram of the SkyProbe data. The peak at 0.065 magnitudes corresponds to the global mode of atmospheric
extinction. To help show this, the distribution of nightly modal attenuations is overplotted as a dashed curve. A linear least-squares fit to the data with 0.40 magnitudes or more gives a power-law slope of -1.84 (thick black line) corresponding to a value of $\alpha=5.43$. See text for details.}
\label{figure_raw}
\end{figure}

Figure~\ref{figure_raw} is a histogram of the SkyProbe data, plotted as attenuation in magnitudes. There is a 
strong peak at 0.065 magnitudes, indicating the global mode of atmospheric attenuation. To help see this, the distribution of nightly modal attenuations is overplotted as a dashed line. The second ``bump" near 0.6 magnitudes is an instrumental artifact. It is caused by untracked exposures at the beginning of the 
night. 
At this time the telescope 
is sometimes parked at zenith and the resulting star trails are not processed properly by the automated software. 
A total of 145 nights identified with the resulting 
bogus zeropoint have been isolated and are not included in the analysis, although it is possible that some untracked exposures remain. 

Away from the peak associated
with the zeropoint, a power-law behavior dominates. A linear least-squares fit to data with 
between 0.4 and 1 magnitude of extinction (solid line; extended above and below the fitted region) gives a power-law slope of -1.84. Comparing this with equation 6 gives $\alpha = -1/({\rm slope}\times A_{\rm 1, cloud}) = -1/(-1.84\times0.10) = 5.43$.
Note how well this curve fits - with constant $\alpha$ - even for attenuations approaching 3 magnitudes. We are not interested in this end of the distribution of course, as there is so little power there, but it is still worth mentioning. A bias against observing high $A$ values - closing the dome for clouds, or systematically pointing the telescope to clearer parts of the sky - would presumably steepen the slope of this curve, perhaps even resulting in a knee at an attenuation where clouds become obvious, say, at 0.5 mag. A bend in the opposite sense would indicate observations taken {\it preferentially} in cloudy conditions. A more plausible scenario, consistent with the CFHT nightly log is that the dome is generally opened for nights that begin as mostly clear (with perhaps some cirrus), and observations continue as long as interruptions by thick cloud are brief. The dome is not opened during poor nights, as this leads to inefficient observing, and instead these nights are allocated to engineering tasks, if possible.

The fraction of power under the curve (lightly shaded region) compared to that in the total distribution is 18.6\%.  This is the total fraction of dome-open time that was cloudy. 
This indicates that for 81.4\% of the SkyProbe samples, attenuation was due solely to air.
SkyProbe was essentially always operational during the period discussed here and took data 
whenever the dome shutter was open. 
Taking this to be 68.8\% of the time, conditions were photometric $0.814\times0.688=56.0$\% of the time. 
The true value must be something more than this because downtime due
to non-weather reasons (when the sky may not be cloudy) have been included, lumped in with times when the dome was closed (darkly shaded region). For example, because SkyProbe is attached to the telescope, 
datataking is subject to the telescope slewing to a new target.

\begin{figure}
\plotonesmall{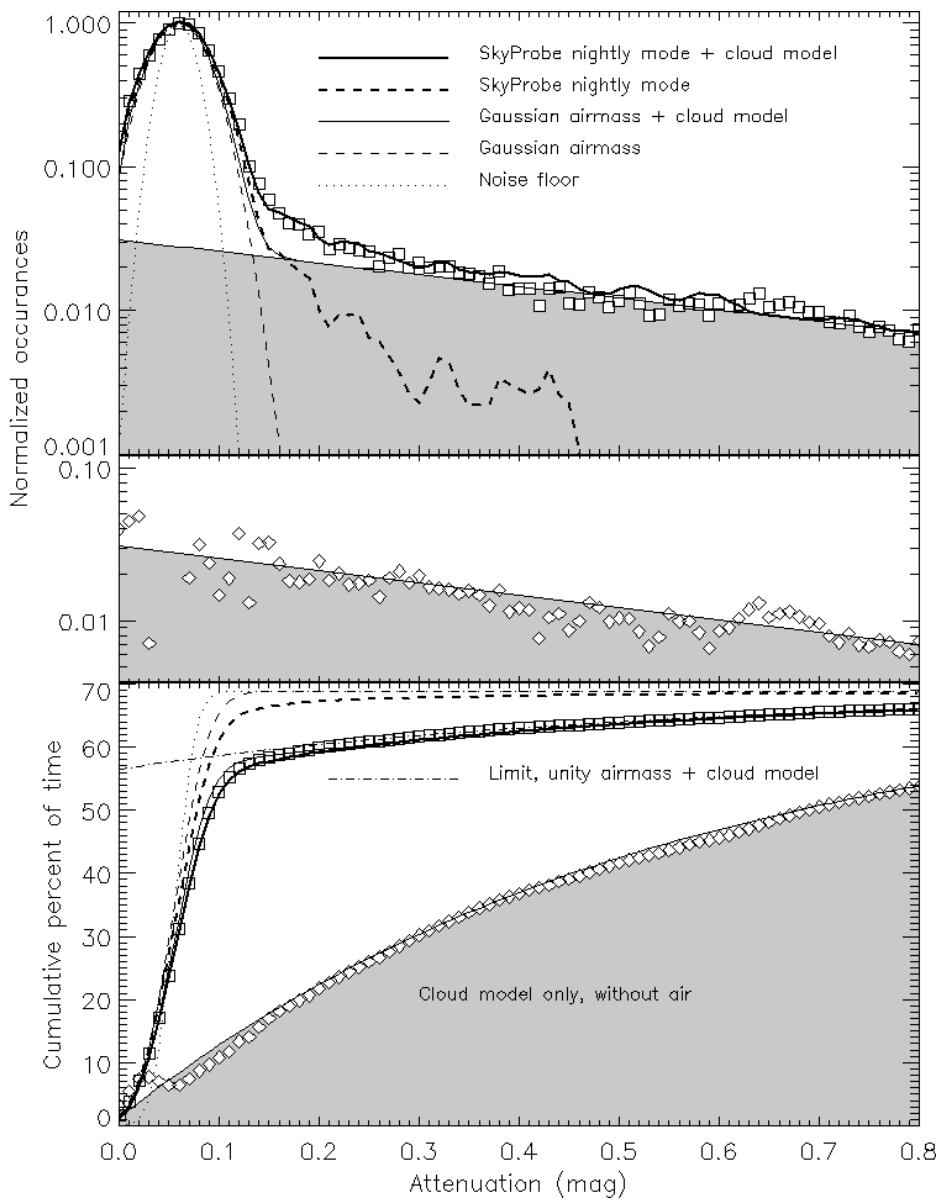}
\caption{A histogram of the data (top panel; squares) that with nightly modal attenuation subtracted (middle panel; diamonds) and cumulative histogram (bottom panel) with a power-law model for clouds indicated by the shaded region. The two models for atmospheric attenuation are plotted as dashed curves, and the combination of these with the cloud model as solid curves. The dotted line indicates the minimum distribution of attenuations for $Z=1$ and photometric scatter of 0.01 mag. The fraction of time during the night without cloud is 56.0\%. This is evidently not strongly dependent on the model of atmospheric attenuation used, as both provide good fits to the data.}
\label{figure_histogram}
\end{figure}

A more detailed histogram is shown in 
Figure~\ref{figure_histogram}, now plotted only for attenuations less than 0.8 magnitudes; the middle panel is that same, except with the nightly modal attenuation subtracted; and the bottom panel is a cumulative histogram of separated air and cloud.
The thick black curve is the result of combining the distribution of nightly modal attenuations (thick dashed curve) with that for the cloud model, indicated by the shaded region. This provides a good fit, although it seems that the tail of the nightly modal distribution includes some nights when thin cloud was prevalent. To see this, the cloud model combined with the simple Gaussian model of atmospheric attenuation is indicated by the thin black curve.
The dotted line indicates Gaussian noise of 0.01 magnitudes about a fixed atmospheric attenuation of 0.065 mag, that is, $\sigma=0$ and no cloud. It is the addition of this noise which sufficiently broadens the distribution of atmospheric attenuation that it leads to the nonphysical result of attenuations approaching zero, even becoming negative.
This effectively restricts the accuracy of subtraction of the modal attenuation - only a spike could be perfectly subtracted - which is the cause of the scatter in the cloud measurements below about 0.2 mag, and ultimately makes the estimate of the clear fraction uncertain to within perhaps 2\%.

\begin{figure}
\plotone{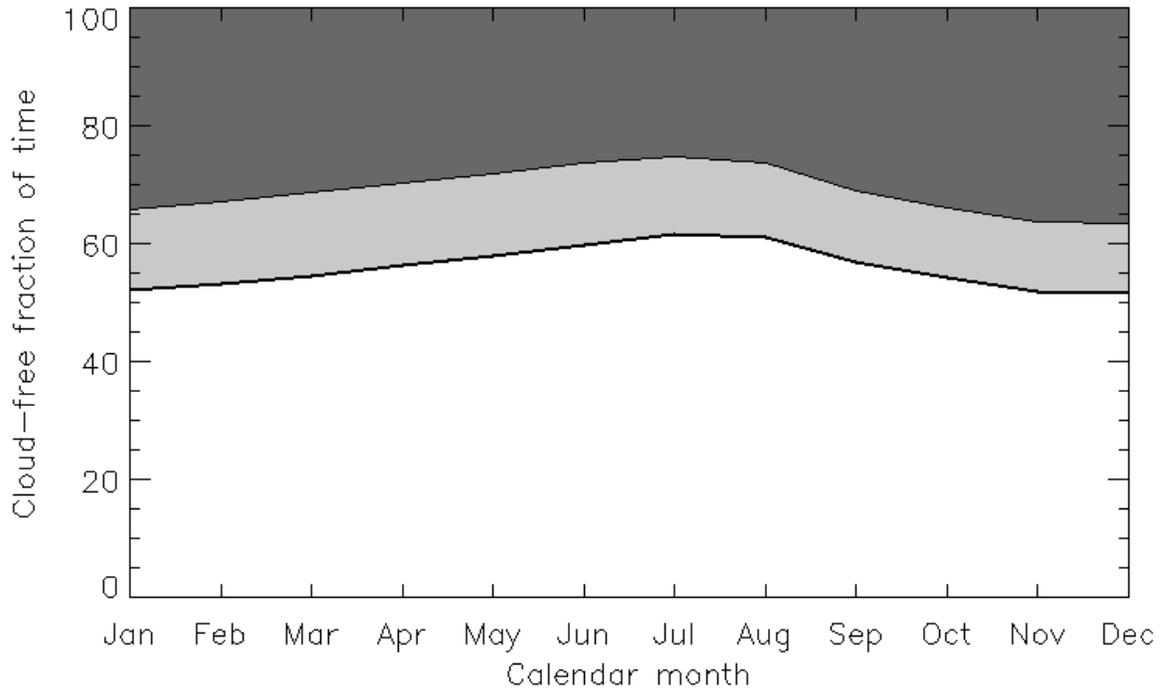}
\caption{A plot of cloud-free fraction of skies for the SkyProbe data calculated for each calendar month (thick curve); lightly shaded region is cloudy. This is corrected for the unobserved fraction of time during the night, indicated by the darkly shaded region. All unobserved samples are assumed to be cloudy.}
\label{figure_monthly}
\end{figure}

To see if any more information can be gleaned from the distribution of SkyProbe attenuations, for example, if the clear fraction of time varies seasonally, this calcuation was repeated after binning the data by calendar month. The results are presented in Figure~\ref{figure_monthly}. The solid line is the calculation of cloud-free fraction of time, with the lightly-shaded region indicating cloud; dark shading indicates the dome is closed. Note that CFHT is very efficient in maximimizing the amount of time that the dome is open when skies are clear; this fraction of time varies moderately over the year, with perhaps a peak in late summer approaching 60\%.
This and the values of other parameters appearing in equations 1 through 6 are given in Table~\ref{table_parameters}.

\subsection{Simulated SkyProbe Data}

So far, we have only fit a power law to the SkyProbe data as per equation 6 without need for an explicit assumption about the duration of cloud-free skies other than to bin the data by month. 
It is interesting to consider for what values of $\delta_0$ equation 10 is a good fit to the data.
A computer simulation was developed to help explore this.
Seven artificial SkyProbe datasets, assuming $\delta_0=$ 1, 2, 6, 12, 24, 72 and 168 hours, were generated as follows: An initial cloud duration was selected at random from between zero and $\delta_0$. If a duration of zero was selected, then a period of $\delta_0$ had no attenuation. Otherwise, $T$ was calculated via equation 10. A power-law slope of $\alpha=5.43$ was used.
In one-minute timesteps until the end of this period all datapoints each had this 
value. Then a new random duration was selected and the process repeated for a number of samples
similar to the observations. Poisson noise of 0.01 magnitudes was added to simulate instrumental uncertainty.
Data lost due to the dome being shut and to overheads was simulated by masking out 31.2\% of nights with the highest median attenuation, just as CFHT closes for cloudy nights. As can be expected, since the mean of $T$ has been defined to be unity, each of these simulations has the same mean attenuation, and correspondingly the same mean duration of clouds, that is ${\bar A}_{\rm cloud}=0.10$ magnitudes and ${\bar \delta}=2.14$ hours. 

\begin{figure}
\plotone{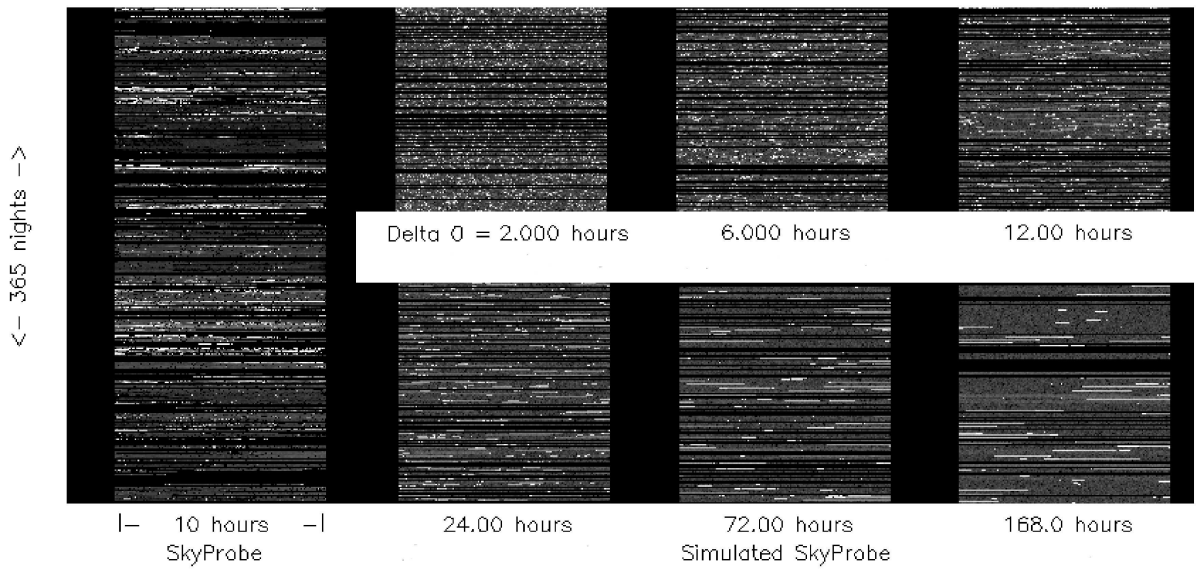}
\caption{Images of the SkyProbe data and simulations. Each pixel represents the average of three SkyProbe samples; white indicating high attenuation with the same stretch used throughout. Noise is included, so clear periods appear grey; times when the dome is closed are black. One night comprises a row. A year of data is shown. Simulations are shown for $\delta_0$ of 2, 6, 12, 24, 72, and 168 hours. None reproduces both the many short thick clouds seen in the data as well as longer ``clouded-out" periods of several days.}
\label{figure_images}
\end{figure}

The results are shown in Figure~\ref{figure_images} along with the SkyProbe data, displayed as 
grey-scale images. The nightly modal attenuation has been subtracted from the data to provide an estimate of attenuation due only to cloud.
White indicates high attenuation; the same stretch is used throughout. Each pixel represents the average of three SkyProbe samples. Each row is one night of data, with a full year along a column. Instrumental noise is included, so clear skies appear as a uniform grey; black indicates the dome is shut.
Small values of $\delta_0$, less than 24 hours, seem to reproduce the short, thick clouds seen in the data. Larger values of $\delta_0$ provide longer ``clouded out" periods, some lasting several days, also evident in the data. But larger values of $\delta_0$ produce too few short, thick clouds. 

\begin{figure}
\plotone{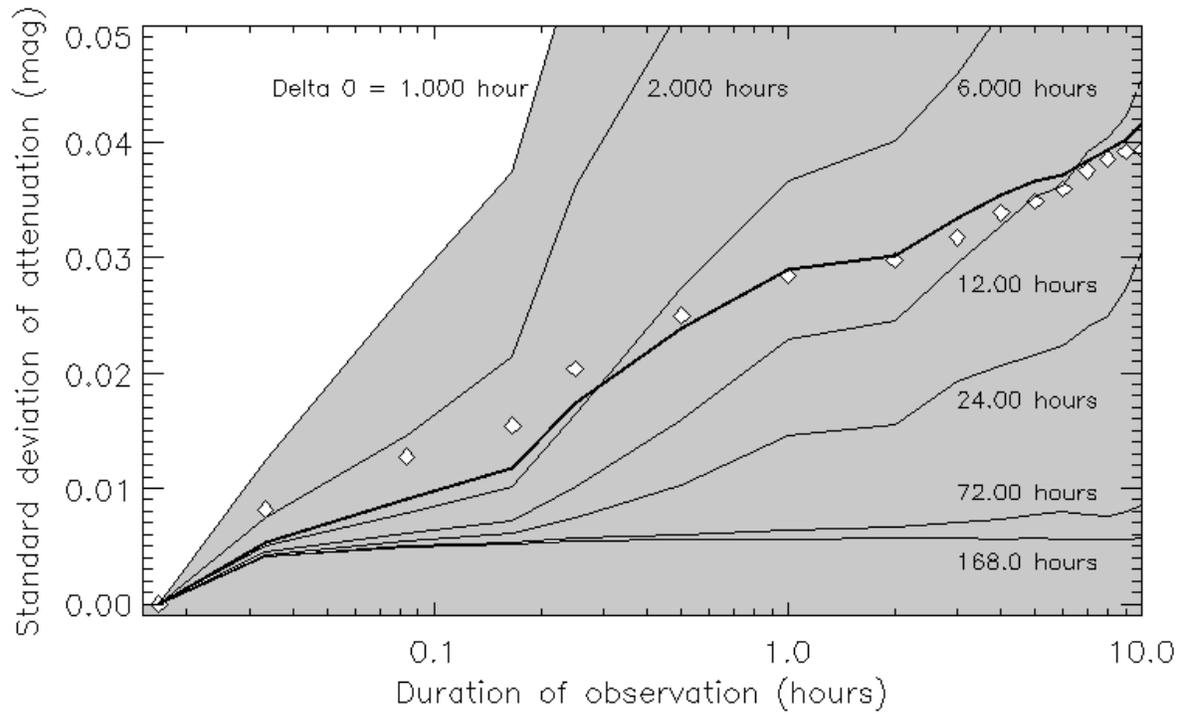}
\caption{A plot of standard deviations in SkyProbe attenuation due to cloud for observations lasting various time intervals, from 1 minute up to 10 hours. The nightly mode of attenuation has been subtracted for each night. The results of the simulations are shown as thin black lines, for $\delta_0$ of 1, 2, 6, 12, 24, 72 and 168 hours; the average of those is indicated by a thick black curve. It would seem a model incorporating an appropriate distribution of $\delta_0$ could reproduce the variation seen in the data.}
\label{figure_statistics}
\end{figure}

It may be that a single value of $\delta_0$ is insufficient to reproduce both long-term periods of cloud and the appropriate variation in attenuation during a night.
This can be seen in Figure~\ref{figure_statistics}, which is a plot of the standard deviation of attenuation as a function of various time intervals. This has been calculated by dividing up each night's SkyProbe data into blocks of a given interval, and calculating the standard deviation over each. This analysis is restricted to attenuations of 0.5 magnitudes or less, so conditions excluding obvious clouds. Shown are values for 1, 2, 5, 10, 15, 30, and 60 minutes, and then every hour up to 10 hours. Note the strong power-law behaviour, as expected from the cloud model. Thin black curves indicate the results for choices of $\delta_0$ shown in Figure~\ref{figure_images} plus that of $\delta_0=1$ hour; the thick black curve is the average of these. Although the average of the models does not exactly reproduce the data, it is a reasonably good fit, especially for observational durations longer than an hour. That the average of the models underestimates variation in attenuation on timescales less than an hour suggests that $\delta_0$ might include shorter timescales than this. And so it seems that a refined distribution of $\delta_0$ - possibly spread over all timescales - could produce a better fit. But determining this distribution is beyond the scope of this paper. One path to it may be to investigate similarities with the Fried parameter $r_0$ associated with optical turbulence, where strong self-similar behaviour is also evident.

A final result, evident in Figure~\ref{figure_statistics}, is that thin clouds set a threshold for photometric accuracy. Even if atmospheric attenuation is subtracted perfectly, if clouds less than 0.5 mag in attenuation are not accounted for, observations lasting an hour or more will typically incur a 0.03 magnitude absolute error in photometry. This ``cloud-induced'' error is comparable to the standard deviation in all MegaCam zeropoints (0.031 mag). As one might expect, the error component due to thin cloud could be reduced by confining observations to better sky conditions.  It is not shown in Figure~\ref{figure_statistics}, but further restricting the SkyProbe analysis to attenuations less than 0.2 magnitudes reduces the slope of the cloud-induced error, yielding a 0.01 mag standard deviation after 1 hour.  Of course, the available observing time under these better conditions is reduced too, down to 60\% based on Table~\ref{table_statistics}; only 56\% of the time will have no clouds at all. This serves as a reminder to observers to obtain frequent photometric calibration; further validation of SkyProbe as an instrument dedicated to measuring sky transparency.

\section{Summary}\label{summary}

We have presented an analysis of CFHT SkyProbe data covering almost three calendar years beginning 
in 2003.
The results are in agreement with MegaCam data during the same period.
The SkyProbe data show that the V-band attenuation had a modal value of 0.065 magnitudes corresponding
to 56\% of time with attenuation due to an atmosphere without clouds, and 
that 59\% of the time V-band attenuation was 0.20 magnitudes or less (including both air and thin cloud). These estimates are conservative, as they include small telescope overheads. Some modest seasonal variation, $\sim 4$\%, is seen in the fraction of time that skies are clear, but ultimately the accuracy in that estimate is perhaps $\pm 2$\%, limited by the photometric uncertainty of SkyProbe.
As these results are for open dome shutter time on CFHT, and account for overheads, they can be used as inputs to 
designing AO with lasers, for example.

A power-law model of cloud attenuation and duration has been presented which is a good fit to the SkyProbe data. It is consistent both with the mean attenuation due to thin clouds, and the appearance of brief thick clouds at times causing the ``clouding out" of nights. It is hoped that this simple model will lead to further investigation of the self-similar nature of clouds on Mauna Kea and elsewhere, which could possibly lead to improvements in photometric accuracy at the telescopes.

\acknowledgements

We thank
Derrick Salmon and Konstantinos Vogiatzis for their assistance in obtaining data from the CFHT weather 
tower and Chris Pritchet for his MegaCam
zeropoints, and thoughtful comments. 
We would like to thank Glen Herriot and Jean-Pierre Veran for helpful 
conversations, and the Thirty Meter Telescope Site-Testing Advisory Group for useful comments on an earlier internal report on these results. This work includes archival data obtained from the National Oceanic and Atmospheric Administration, Climate Monitoring and Diagnostics Laboratory, Aerosols Group.

\clearpage

\begin{deluxetable}{lcccccc}
\tablecaption{Breakdown of SkyProbe and MegaCam Observing Statistics in Nights per Year\tablenotemark{1}\label{table_statistics}}
\tablewidth{0pt}
\tabletypesize{\small}
\tablehead{&\multicolumn{3}{c}{Unobserved} & &\multicolumn{2}{c}{Attenuation (mag)}\\
\cline{2-4} \cline{6-7}
\colhead{Instrument} &\colhead{Dome shut} &\colhead{Overheads\tablenotemark{2}} &\colhead{Total} &\colhead{Observed} &\colhead{$\ge0.2$} &\colhead{$<0.2$}}
\startdata
SkyProbe  &87 (23.8)  &27 ~(7.4)  &114 (31.2)  &251 (68.8)  &35 (9.6)  &216 (59.2)\\
MegaCam   &87 (23.8)  &37 (10.2)  &124 (34.0)  &241 (66.0)  &20 (5.5)  &221 (60.5)
\enddata
\tablenotetext{1}{Averages for a total sample of 1035 nights with SkyProbe and 541 nights with MegaCam. Fractional nights have been rounded down. Percent of total nights in the respective samples are shown in brackets.}
\tablenotetext{2}{Includes time lost during telescope slews, detector readout, technical faults, and engineering downtime.}
\end{deluxetable}

\clearpage

\begin{deluxetable}{lcccc}
\tablecaption{Atmospheric Parameters Derived from SkyProbe\tablenotemark{1}\label{table_parameters}}
\tablewidth{0pt}
\tabletypesize{\small}
\tablehead{\colhead{$A_{\rm air,1}$} & &\colhead{$A_{\rm cloud,1}$} & &\colhead{$f_{\rm clear}$}\\
\colhead{(mag)} &\colhead{$\sigma$} &\colhead{(mag)} &\colhead{$\alpha$} &\colhead{(percent)}}
\startdata
$0.065\pm{0.002}$ &$0.027\pm{0.002}$ &$0.095\pm{0.003}$ &$5.43\pm{0.13}$ &$56.0\pm{3.6}$\\
\enddata
\tablenotetext{1}{Averages after binning by calendar month. Quoted uncertainties are the standard deviations in each parameter.}
\end{deluxetable}

\clearpage

\clearpage

\clearpage

\clearpage

\clearpage

\clearpage

\end{document}